\documentclass[12pt,epsfig,epsf]{article}
\usepackage[T1]{fontenc}

\makeatletter

\newcommand{\LyX}{L\kern-.1667em\lower.25em\hbox{Y}\kern-.125emX\@}

\makeatother

\begin{document}

\newcommand{\nn}{\noindent} \renewcommand{\thefootnote}{\fnsymbol{footnote}}

{\par\centering \textbf{\large The quantum anharmonic oscillator in the Heisenberg
   picture and multiple scale techniques }\large \par}

{\par\centering \vspace{.4cm} {\large G. Auberson and M. Capdequi Peyran\`{e}re}\large \par}

{\par\centering {\large \vspace{.6cm} Laboratoire de Physique Math\'{e}matique
et Th\'{e}orique, CNRS-UMR 5825}\\
 {\large Universit\'{e} Montpellier II, F--34095 Montpellier Cedex 5, France }\large \par}

\vspace{1.2cm}
{\par\centering \vspace{.4cm} {\large PM 0136 }\large \par}

\vspace{1.2cm}
\nn Multiple scale techniques are well-known in classical mechanics to give
perturbation series free from resonant terms. When applied to the quantum anharmonic
oscillator, these techniques lead to interesting features concerning the solution
of the Heisenberg equations of motion and the Hamiltonian spectrum.

\vfill

\newpage

\pagestyle{plain} \renewcommand{\thefootnote}{\arabic{footnote} } \setcounter{footnote}{0}

\section*{0.~Introduction}

Multiple scale techniques (MST) originated in Poincar\'{e} works have
been developed by many authors, mainly in solving (partial) differential equations
related to physical problems in celestial mechanics or in fluid dynamics. All
these methods have a common mathematical purpose: to
avoid resonances or secularities appearing in the usual or conventional perturbative
theory. From a more physical point of view, one can see the MST as adaptable
methods that feel the underlying physical phenomena in order to fit them. In
other words, the usual perturbative theory tends to impose its choices while
MST are flexible and compose with the real medium.

In this work, we apply one of the various MST to the quantum anharmonic oscillator.
Such studies have been initiated by Bender and Bettencourt (B\&B) in two recent
papers \cite{BB1,BB2}. They have found that the non resonance condition leads
to a \char`\"{}mass renormalisation\char`\"{} of the oscillator and 
- as a by-product - to the energy level differences of the quantum oscillator.
This pioneering work was limited to the first non trivial order in MST perturbation
of the coupling constant of the anharmonicity. The aim of the present paper is to extend
this early study in several directions. First, we introduce an alternative framework,
 which turns out to be more convenient than the B\&B one for performing higher order
 calculations. Secondly, it turns out that we are able to obtain the
energy levels themselves at these perturbative orders. In the third point, we show that
the diagonalization of the Hamiltonian is rather easy once the free Hamiltonian has been
recast in an appropriate form. Finally, this approach leads to a natural
and elegant method to find perturbatively the eigenvalues of the full Hamiltonian,
far away from the original MST concept.

The paper is organized as follows. In the first section, although the classical
anharmonic oscillator is studied in details in many textbooks \cite{JK,N},
we sketch some relevant points in order to further clarify the differences and
the analogies between the classical and the quantum cases. In the second section
we explain our framework and we work out the two first orders in MST
perturbation, that includes the full solution of the Heisenberg equations and
the energy levels. The third section is devoted to general arguments showing that the
method is compelled to work at any order, due to its connection with a certain unitary
transformation which diagonalizes the Hamiltonian of the anharmonic oscillator.
We postpone to an appendix some explicit results: the solutions of the Heisenberg
equations of motion and the energy levels of the full Hamiltonian, up to the
order 6 included. .

\section*{1.~The Classical case}

The classical anharmonic oscillator (CAO) is probably one of the most popular
examples where the conventional and MST perturbative theories lead to obvious
differences. Often, one speaks of Duffing equation instead, although
this equation is nothing but the equation of motion of the CAO. To be precise,
the Duffing equation is a second order non linear equation in the time variable,
the solution of which being the position of the CAO. Starting from the CAO Lagrangian
( in units where the mass parameter is 1 ) 
\[
L(q,\dot{q})=\dot{q}^{2}/2-\omega ^{2}q^{2}/2-gq^{4}\]
one readily gets from the Euler-Lagrange equation:

\[
(e):\ddot{q}+\omega ^{2}q+4gq^{3}=0.\]

The usual formal perturbation expansion reads 

\[
q(t)=\sum_{n=0}^{\infty} g^{n}q_{n}(t),\]
 ( with some initial conditions, say \( q(0)=Q \) and \( \dot{q}(0)=0 \)
), and the first equations one obtains from (e) are :

\[
(e_{0}):\ddot{q}_{0}+\omega ^{2}q_{0}=0 ,\]

\[
(e_{1}):\ddot{q}_{1}+\omega ^{2}q_{1}=-4q^{3}_{0} .\]

 Then the frequency of the solution of the homogeneous part of ($e_1$) coincides with
 the frequency of $q_0(t)$ = $Q cos\omega t$, which generates a resonance in the solution
 $q_1(t)$ of the full equation ($e_1$):

\[
q_{1}(t)= \frac{Q^{3}}{8\omega^2} (cos3\omega t - cos\omega t- 12 \omega t sin\omega t).\]

Hence $q_1(t)$ is unbounded and the truncated expansion \( q_{0}(t)+gq_{1}(t) \)
cannot be an acceptable approximation of $q(t)$ for times $t$ larger than $\omega/Q^2 g$,
 however small $g$ may be. The flaw is even worse at the higher orders. It is obvious on this simple example that the perturbative
solution develops spurious behaviour which is absent in the exact solution.
Indeed, it is well known that the exact solution is bounded and periodic.

The main idea of MST for dealing with this problem is the introduction of new variables,
independent and appropriate, and we refer to textbooks for an extensive review
of the various possibilities. Here we concentrate on the anharmonic oscillator.
Some methods take into account ab initio that the circular functions play a
major role. For instance, in the Poincar\'{e} method, one looks for sine and
cosine solutions whose argument is still \( \omega t \) but where \( \omega  \)
is now an arbitrary function of the coupling constant, actually \( \omega =\Sigma g^{n}\omega _{n} \).
Then one has to find the \( \omega _{n} \)'s , order by order, to discard the resonance.
We do not insist on the application of these methods to the CAO because we believe they 
are not
suitable for the quantum case. Another class of MST seems to be of a larger use, since
there is no \char`\"{}prerequisite\char`\"{} in these methods. The MST we will
use, also called Derivative Expansion Method, belongs to this class: it promotes
the time variable to be a function of the coupling constant, namely \( t_{n}=g^{n}t. \)
Actually, the method is not so rough and one first extends the function depending
on \( t \) to an \char`\"{}extended\char`\"{} function depending on all the 
variables \( t_{n} \) ($n=0,1,2,...$) assumed to be independent \cite{S}. So, one introduces
a new position function \( Q(T,g) \) depending on the collection \( T=\{t_{0},t_{1},t_{2},...\} \)
of independent variables \( t_{n} \). This function is considered as an extension
of the true position in the Lagrange formalism, which is recovered by restricting
\( Q \) to the section \( t_{n}=g^{n}t \) of the \( T- \)space: $q(t,g)=Q(T,g)|_{t_{n}=g^{n}t}$.

Then, forgetting temporarily any reference to the coupling constant in these
\( t_{n} \) variables, one expands in power of \( g \) the new position
function \( Q \) : 
\[
Q(T,g)=\sum_{n=0}^{\infty} g^{n}Q_{n}(T).\]

One obtains from ($e$) the following set of equations, limited here at the three first
orders :

\[
D_{0}^{2}Q_{0}(T)+\omega ^{2}Q_{0}(T)=0\, ,\]

\[
D_{0}^{2}Q_{1}(T)+\omega ^{2}Q_{1}(T)=-2D_{0}D_{1}Q_{0}(T)-4Q^{3}_{0}(T)\, ,\]

\[
D_{0}^{2}Q_{2}(T)+\omega ^{2}Q_{2}(T)=-(D_{1}^{2}+2D_{0}D_{2})Q_{0}(T)-2D_{0}D_{1}Q_{1}(T)
-12Q^{2}_{0}(T)Q_{1}(T),\]

 using \( d/dt=\Sigma _{n}g^{n}D_{n} \) , \( D_{n}=\partial /\partial t_{n} \)
. 

The basic principle of the method now consists in adjusting the \( t_{1} \) dependence of
$Q(T)$ so as to eliminate the secularity in the second equation, next the \( t_{2} \)
 dependence of $Q(T)$ so as to eliminate the secularity in the third equation, and so on. We 
shall not work out the derivation here
(it can be found for example in ref 3 or 4) and we merely give the solution up
to the second order in \( g \) in its final form, for further classical versus quantum discussions
:

\[
q(t,g)=\frac{a}{2} [\exp(-i(\Omega t+b))+
   \frac{\lambda}{8} (1- \frac{21 \lambda}{8}) \exp(-3i(\Omega t+b)) + 
 \frac{\lambda^2}{64}
        \exp(-5i(\Omega t+b))] + C.C.,\]
where :

\begin{equation}
\label{1}
\Omega =\omega (1+\frac{3\lambda}{2} -\frac{15 \lambda^2}{16} ), 
\lambda = \frac{g a^2}{\omega^2}
\end{equation}
 and a and b are two real integration constants fixed by the initial conditions (here unspecified).

Since the (perturbative) energy is conserved, it can be computed most easily by choosing 
 \(t=-b/\Omega \) or \( t=(\frac{\pi}{2}-b)/\Omega \) in $q(t,g)$ :

\begin{equation}
\label{2}
E_{c}=\frac{a^{2}\omega ^{2}}{2} (1  +\frac{9 \lambda}{4} +
\frac{25 \lambda^{2}}{64}) +O(\lambda^{3}).
\end{equation}

We conclude this first section by a few comments. As far as we know, all the multiple
scale techniques dealing with the secularities of the classical anharmonic oscillator
are successful. However this is not a general feature, and some methods are
not suitable for certain problems. Moreover it is absolutely not our purpose to discuss
on a rigorous basis the mathematical aspects of the secular or non secular perturbative
 expansions.

\section*{2.~The Quantum case : Derivation}

The quantum anharmonic oscillator (QAO) has been studied in the paper of B\&B
through the Heisenberg equation of motions for the relevant operators and we will follow
this method. The main difference
between the work of B\&B and ours is that we will use the creation and annihilation
operators to manage the problem of removing the secularities. At first sight
the gain in doing this choice is not obvious and perhaps not essential. Moreover
one can detect in the B\&B paper an indication pointing to this direction.
Let us look at the couple of equations (21) in their work \cite{BB1}, which can
be written as :

\[
D_{1}Y=-CX-XC\, \, \textrm{and}\, \, D_{1}X=CY+YC\]
 where \( X \) and \( Y \) are \( t_{1} \) dependent, self-adjoint operators while \( C \)
is a constant, self-adjoint operator. The authors proceed with some arguments
 \char`\"{}suggesting\char`\"{}
the form of the solution, with the help of Weyl ordered products and Euler polynomials
to deal with these equations. Of course, it seems difficult, or at least hazardous,
to generalize at high orders a \char`\"{}suggestive\char`\"{} method, which could
be seen as a reminiscence of the Poincar\'{e} method, but we have more convincing
arguments to leave this path. First, using $Z = X + i Y$, the previous couple of equations
reduces to the single equation :   
\[
D_{1}Z(t_{1})=-i(Z(t_{1})C+CZ(t_{1}))\]
 whose solution is \( Z(t_{1})=\exp(-iCt_{1})Z(0)\exp(-iCt_{1}) \), as it is
easy to check. The operator \( Z(t_{1}) \)
is closely related to the creation/annihilation operators. Once derived the expression
 of the creation/annihilation operators, it is not necessary, in order to 
proceed further,
to write down the position operator. Indeed, almost all the informations, the
\char`\"{}mass renormalization\char`\"{} effect and the difference of energy
levels, are already contained in the argument of the exponentials. Secondly,
the Heisenberg equations in terms of creation/annihilation operators are first
order differential equations in place of the second order one for the
position operator, which simplifies noticeabily the whole procedure. To be honest,
one has the disadvantage to carry both creator and annihilator, but this is
not a serious complication. Lastly, there appears also a large variation
between the B\&B works and ours in the status of the initial conditions: we
do not use these conditions as in the classical case, which is the way taken
by B\&B. This point will become obvious throughout our study. 

We start with the QAO Hamiltonian \( H \) written in terms of the momentum \( p \)
and position \( q \) operators in convenient units \( (\hbar =\omega =1) \) : \( H=p^{2}/2+q^{2}/2+gq^{4} \),
where \( g \) is assumed to be a \char`\"{}small\char`\"{} (positive) coupling
constant. Whithin the Heisenberg picture, the dynamics is governed by the equations :

\[
 \dot{q}= i[H,q]\, ,\, \dot{p}= i[H,p],\]

supplemented by the canonical commutation relation $ [q,p]=i $, valid at all times. 
The Heisenberg equations give : \( \dot{q}=p \) and \( \dot{p}=-q-4gq^{3} \).
Writing as usual \( q=(a+a^{\dag })/\sqrt{2} \) and \( p=-i(a-a^{\dag })/\sqrt{2} \),
the Hamiltonian becomes:
 
\begin{equation}
\label{3}
H(a,a^{\dag },g)=1/2+a^{\dag }a+g(a+a^{\dag })^{4}/4
\end{equation}
 together with  : 
\begin{equation}
\label{4}
[a(t,g),\, a(t,g)]=1\, ,\, \, \forall t,
\end{equation}
where, to avoid possible confusion later on, we have kept track of the variables $t$ and $g$.

The Heisenberg equation for the annihilator : 

\[
\dot{a}(t,g)=i[H(a(t,g),a^{\dag }(t,g),g),a(t,g)],\]
 reads, in our case :

\begin{equation}
\label{5}
\dot{a}(t,g)=-i(a(t,g)+g(a(t,g)+a^{\dag }(t,g))^{3}).
\end{equation}

Since the Hamiltonian is conserved, its formal solution is:

\[
a(t,g)=\exp(iH(a(0),a^{\dag }(0),g)t)a(0)\exp(-iH(a(0),a^{\dag }(0),g)t),\]
 with \( a(0)\equiv a(0,g) \).

We now turn on the formal series of the multitime perturbative expansion,
similar to that used in the classical case. First one introduces an operator valued
function \( A(T,g) \) depending on the collection \( T \) of independent variables
\( t_{j} \). This function is considered as an extension of the true annihilation
operator in the Heisenberg picture, which is recovered through the restriction:
\begin{equation}
\label{6}
a(t,g)=A(T,g)|_{t_{j}=g^{j}t}. 
\end{equation}

Then the time derivative becomes :

\[
\dot{a}(t,g)=\sum _{n\geq 0}g^{n}D_{n}A(T,g)|_{t_{j}=g^{j}t} .\]

Secondly, \( A(T,g) \) is expanded as :

\begin{equation}
\label{7}
A(T,g)=\sum _{n\geq 0}g^{n}A_{n}(T).
\end{equation}

As for the initial conditions to be associated with the equation of motion (5),
one notices that 
\begin{equation}
\label{8}
a(0,g)=\sum _{n\geq 0}g^{n}A_{n}(0).
\end{equation}

This forces us to choose between two possible starting viewpoints : \\
either a) : \( a(0,g) \) is taken as independent of \( g \) , which implies
\begin{equation}
\label{9}
A_{n}(0)=0,\forall n\geq 1,
\end{equation}
 or b) : the previous condition is not imposed, in which case the initial values
of \( a(t,g) \) must be considered as a function of \( g \). 

It turns out that both approaches lead to consistent multitime expansions.
In fact, the choice a) was (implicitely) adopted by B\&B.
However, these authors did not extend their analysis beyond the first order. In this paper,
we rather follow the procedure b), which we found much more convenient, and in a
sense, more natural. 

The equation of motion for \( a(t,g) \) gives us the following infinite system
for the \( A_{n}(T) \)'s :

\begin{equation}
\label{10}
D_{0}A_{n}+iA_{n}=-\sum ^{n-1}_{m=0}D_{n-m}A_{m}-i\sum _{m,r,s,>,0\atop m+r+s=n-1}Q_{m}Q_{r}Q_{s}\, \, \, \, (n=0,1,2..)
\end{equation}
 where \( Q_{n} \) =\( A_{n}+A^{\dag }_{n} \), or explicitely :\\

\( D_{0}A_{0}+iA_{0}=0, \) \hfill (10.a)

\( D_{0}A_{1}+iA_{1}=-D_{1}A_{0}-iQ_{0}^{3}, \) \hfill (10.b)

\( D_{0}A_{2}+iA_{2}=-(D_{2}A_{0}+D_{1}A_{1})-i(Q_{0}^{2}Q_{1}+Q_{0}Q_{1}Q_{0}+Q_{1}Q_{0}^{2}), \)
\hfill (10.c)\\
 etc...

A simple check shows us that {\bf any} formal solution of (10) generates via (6) and
(7) a formal solution \( a(t,g) \) of (5). In particular, this implies that,
for such a solution, \( [A(T,g),A^{\dag }(T,g)]|_{t_{j}=g^{j}t} \)
is independent of \( t \). Of course, this does no mean yet that \( [A(T,g),A^{\dag }(T,g)] \)
is independent of \( T \), allowing us to impose : 
\begin{equation}
\label{11}
[A(T,g),A^{\dag }(T,g)]=1\, ,\, \forall T\, ,
\end{equation}
 in order to insure the canonical commutation relation (4) . However, one can
look for {\bf those} solutions of (10) which are subjected to the stronger condition
(11), if such solutions do exist indeed, i.e. if no inconsistencies or obstructions
arise in their iterative construction. Together with (7), this entails : 

\begin{equation}
\label{12}
\left\{ \begin{array}{ll}
\displaystyle [A_{0}(T),A^{\dag }_{0}(T)]=1 & \\
 & \forall T\\
\sum _{m=0}^{n}[A_{m}(T),A^{\dag }_{n-m}(T)]=0,\, \, n\geq 1 & 
\end{array}\right. 
\end{equation}

We are now ready to construct step by step the resonance-free
solution of the problem. To zeroth order, the equation (10.a) and the first
 equation (12) yield :

\begin{equation}
\label{13}
A_{0}(T)=A_{01}(T_{1})\exp(-it_{0})
\end{equation}
 with 
\begin{equation}
\label{14}
[A_{01}(T_{1}),A^{\dag }_{01}(T_{1})]=1,\, \forall T_{1},
\end{equation}
 and the notation : \( T_{k}=\{t_{k},t_{k+1},...\}, (k=1,2,...) \).

Then, one can proceed to the first order step by inserting eq (13) into 
eq (10.b) :

\[
D_{0}A_{1}+iA_{1}=-(D_{1}A_{01}+i(A_{01}^{2}A^{\dag }_{01}+A_{01}A^{\dag }_{01}A_{01}+A^{\dag }_{01}A_{01}^{2}))\exp(-it_{0})-i(A_{01}^{3}exp(-3it_{0})\]
 
\begin{equation}
\label{15}
+A^{\dag ^{3}}_{01}\exp(+3it_{0})+(A^{\dag 2}_{01}A_{01}+A^{\dag }_{01}A_{01}A^{\dag }_{01}+A_{01}A^{\dag 2}_{01})exp(+it_{0})).
\end{equation}

Before integrating this equation, one has to get rid of the first
resonant term on the right hand side, which would produce a contribution growing
linearly with \( t_{0}(=t) \). This leads to the condition :

\begin{equation}
\label{16}
D_{1}A_{01}=-i(A_{01}^{2}A^{\dag }_{01}+A_{01}A^{\dag }_{01}A_{01}+A^{\dag }_{01}A_{01}^{2})
\end{equation}
 which will fix the \( t_{1} \) dependence of \( A_{01} \). 

To do that, let us first introduce the self-adjoint operator \( N(T)=A^{\dag }_{0}(T)A_{0}(T) \).
Thanks to (13) and its creator version, \( N(T) \) is only \( T_{1} \) dependent:
\( N(T)=A^{\dag }_{01}(T_{1})A_{01}(T_{1}) \). Moreover as a consequence of
(14), \( A_{01}(T_{1})N(T_{1})=(N(T_{1})+1)A_{01}(T_{1}) \). Lastly, from (16)
, one observes that \( D_{1}N(T_{1})=0 \). Thus \( N \) is also independent
of \( t_{1} \) and (16) can be now written in the tractable form : 

\[
D_{1}A_{01}=-3iA_{01}(T_{1})N(T_{2}),\]
 which produces : 
\[
A_{01}(T_{1})=A_{02}(T_{2})\exp(-3iN(T_{2})t_{1}).\]

This allows us to write down the first order annihilation operator :

\begin{equation}
\label{17}
A_{0}(T)=A_{02}(T_{2})exp(-i(t_{0}+3N(T_{2})t_{1})).
\end{equation}

At the same time, (14) becomes : 
\begin{equation}
\label{18}
[A_{02}(T_{2}),A^{\dag }_{02}(T_{2})]=1,\forall T_{2}.
\end{equation}

One can now come back to the form of (15) exempted of secularity to obtain its
general solution :

\[
A_{1}(T)=A^{3}_{01}(T_{1})\exp(-3it_{0})/2-A_{01}^{\dag 3}(T_{1})exp(+3it_{0})/4-3N(T_{2})A^{\dag }_{01}(T_{1})\exp(+it_{0})/2\]

\begin{equation}
\label{19}
+C_{1}(T_{1})exp(-it_{0}).
\end{equation}
 where the operator \( C_{1}(T_{1}) \) is an integration "constant". The latter
must be so adjusted, if possible, as to insure that the second equation (12) ,

\begin{equation}
\label{20}
[A_{0}(T),A^{\dag }_{1}(T)]+[A_{1}(T),A^{\dag }_{0}(T)]=0,
\end{equation}
 be fulfilled at all times \( T \). Here, it turns out that (20) is
satisfied by taking simply \( C_{1}(T_{1})=0 \). One ends up with :

\begin{equation}
\label{21}
A_{1}(T)=A^{3}_{0}(T)/2-A^{\dag 3}_{0}(T)/4-3N(T_{2})A^{\dag }_{0}(T)/2
\end{equation}
 and the first order step is complete. 

Before going further, some comments are in order. First, writing the position
operator \( q_{0}+gq_{1} \), one notes that the coefficients of \( \exp(\pm it_{0}) \)
 in
\( q_{0} \) get corrections coming from \( q_{1} \). It means, in the position
formalism, the scheme used by B\&B, that one would have to take into
account the solutions of the homogeneous second order differential equation.
Secondly, it appears in (21) that any power of \( exp(+it_{0}) \) (resp. \( \exp(-it_{0}) \))
is multiplied by the same power of \( A^{\dag }_{01}(T_{1}) \) (resp. \(A_{01}(T_{1}) \)).
Such a correspondance, which is specific to our way of managing the initial conditions, will
 be a guide throughout our study. Lastly the solution of the homogeneous equation in the
classical case is different. This variation with the quantum case is due to the different
status of the initial conditions. 

Clearly, one can go iteratively through the higher order steps by similar (although
rapidly tedious) calculations as long as the integration \char`\"{}constants\char`\"{}
analogous to \( C_{1}(T_{1}) \) can be properly adjusted. As in the first
order step, we gather in eq (10.c) the terms containing \( exp(-it_{0}) \),
since \( \exp(-it_{0}) \) is again (and always) solution of the homogeneous equation.
Because \( D_{1}A_{1}(T) \) does not provide such a term, we just have to take
into account the non derivative part of the right hand side of eq (10.c). Through an 
intensive use of the relation $A_{01}(T_1) N(T_2) = (N(T_2) +1) A_{01}(T_1)$, 
this expression can be reduced to:
\( -3A_{02}(T_{2})(17N^{2}(T_{2})+7)\exp(-i t_0 -3iN(T_2)t_1)/4 \),
 and the non resonance
condition coming from the second order reads : 
\[
D_{2}A_{02}(T_{2})= 3iA_{02}(T_{2})(17N^{2}(T_{2})+7)/4.\]
 This equation shows that \( N(T_{2}) \) is in fact
independent of \( t_{2} \), too,(i.e. $N(T_2)=A^{\dag }_{03}(T_{3})A_{03}(T_{3})$ )
 and we find through integration:

\begin{equation}
\label{22}
A_{02}(T_{2})=A_{03}(T_{3})\exp(+3i(17N^{2}(T_{3})+7)t_{2}/4),
\end{equation}
 whereas (18) becomes: 
\begin{equation}
\label{23}
[A_{03}(T_{3}),A^{\dag }_{03}(T_{3})]=1,\, \forall T_{3}.
\end{equation}

Collecting equations (7), (17) and (22), we see that the non resonance conditions, up to the second order, imply that
the first order term of the expansion of the annihilation operator is, in the variable \( t \): 

\begin{equation}
\label{24}
a_{0}(t,g)=a_{0}(0,g)(\exp(-it(1+3gN-3g^{2}(17N^{2}+7)/4))+O(g^{3})),
\end{equation}
 which exhibits a large difference with the classical case : 17 is a prime number,
difficult to link with the other prime number 5 coming in
the CAO frequency (1) .
We will discuss later on this CAO/QAO (apparent) discrepancy. Nevertheless, the result,
equation (24), is in perfect agreement with the perturbative expression of the energy
levels of the QAO, as calculated by standard methods : 

\begin{equation}
\label{25}
E_{n}(g)=1/2+n+3g(1+2n+2n^{2})/4-g^{2}(1+2n)(21+17n+17n^{2})/8+O(g^{3}).
\end{equation}

Indeed, a straightforward argument based on the formal expression of $a(t,g)$ in the 
Heisenberg picture shows us that the frequency appearing in (24) for $N=n$ should coincide
 with
\( E_{n}(g)-E_{n-1}(g) \). This is readily checked. 

Turning back on the second order equation (10.c) cleared from its resonant
terms, we obtain its general solution :

\[
A_{2}(T)=-15A_{0}^{3}(N-1)/4+3A_{0}^{5}/16+3(23N^{2}+7)A^{\dag }_{0}/8\]
 
\begin{equation}
\label{26}
+21(N-1)A_{0}^{\dag 3}/16-A_{0}^{\dag 5}/8+C_{2}(T_{1})\exp(-it_{0}),
\end{equation}
 where \( A_{0} \) and \( N \) stand for \( A_{0}(T) \) and \( N(T_{3}) \). In contrast 
with $C_1(T_1)$ in (19), the operator \( C_{2}(T_{1}) \) cannot be taken as vanishing, because
the second condition (12), 

\[
[A_{0}(T),A^{\dag }_{2}(T)]+[A_{1}(T),A^{\dag }_{1}(T)]+[A_{2}(T),A^{\dag }_{0}(T)]=0,\]
 would not be fulfilled. Imposing this and using eqs (12),(17),(18) and (26),
one finds instead an appropriate expression for the solution of
the homogeneous version of (10.c), namely :

\begin{equation}
\label{27}
C_{2}(T_1)\exp(-it_0)=-9A_{0}(T)(1-3N^{2})/32.
\end{equation}

Let us notice that the (operator) coefficients of \( \exp(\pm it_{0}) \) which
appear in the zeroth order solution get corrections from the first and second orders,
and the coefficients of \( \exp(\pm 3it_{0}) \) which appear at the
first order get also corrections coming from the second order. Such a behaviour
still holds at the third order, as we have checked.

So far, the perturbative expression of the energy levels of the QAO (which was not our main 
goal) did not show up in full within our MST procedure. Yet, it can be found 
(without
 appealing to other perturbative methods) by inserting $a(t,g)$ as given by
 equations (6), (21), (26) and (27) in the Hamiltonian (3). Obviously, we
are waiting for an expansion in powers of g polynomially dependent on  \( A_{0} \) 
and \( A^{\dag }_{0} \),
up to the second order in \( g \) :

\[
H=H_{0}+gH_{1}+g^{2}H_{2}+O(g^{3}),\]

The result is that the \( H_{j} \)'s are function of \( N=A^{\dag }_{0}A_{0} \),
not of \( A_{0} \) and \( A^{\dag }_{0} \) separately : 
\begin{equation}
\label{28}
H=1/2+N+3g(1+2N+2N^{2})/4-g^{2}(1+2N)(21+17N+17N^{2})/8+O(g^{3}).
\end{equation}

 This feature, which technically appears as an accident due to many cancellations,
  is in fact easy to understand.
One observes, at each step, the $t_0$, $t_1$, $t_2$... dependences of $A_0(T)$
arise from the exponentials only. Since the $t$-dependence must eventually disappear
from the conserved quantity $H$, a proper balance between $A_0(T)$ and $A^{\dag}_0(T)$ 
is expected in each of the monomials $H_j$, namely as many creators as annihilators.
Then, whatever are the number and the order of the $A_0$'s and $A^{\dag}_0$'s in those
polynomials, the commutatiom relation (12) allows us to cast the $H_j$'s in the form of 
polynomials in $N= A^{\dag}_0 A_0$. Furthermore, anticipating a result to be proved in the
next section, $N(T)$ is not only independent of $t_0$, $t_1$ and $t_2$ but in fact of $T$
altogether: $N=A^{\dag}_0(0) A_0(0)$. On account of the Heisenberg algebra (12) (taken at 
$T=0$) this implies that the spectrum of $N$ is the set of non negative integers, and 
the expression (28) for $H$ is in complete agreement with (25) indeed.
 
Finally, we have to explain the apparent discrepancy noticed earlier between the 
classical
and quantum results. Let us write the classical energy (2) for \( \omega =1 \):
\ \( E_{c}=a^{2}/2+3ga^{4}/8+O(g^2) \). Now for large \( n \) the quantum energy (25) 
reduces to : 
 $E_q = n+3gn^2/2+O(g^2)$.
Then the natural correspondance is \( a^{2}/2\rightarrow n \) plus a quantum correction
so adjusted as to insure \( E_{c} = E_{q}+O(g^2) \). 
One finds \( a^{2}/2\rightarrow n - 3gn^2\) which, inserted
in the classical frequency (1) gives \( \Omega =1+3gn-51g^{2}n^2/4 \), in agreement 
with the large $n$ quantum frequency from (24), derived within the MST scheme.

\section*{3.~The Quantum case : General discussion}

In the previous section, in particular on eqs (21),(26) and (27), one observes
that the construction is made with two elementary bricks \( A_{0}(T) \)
and \( A^{\dag }_{0}(T) \), where \( A_{0}(T) \) is the first term in the MST
expansion of the annihilation operator. We have also pointed out the simple connection 
between the operator
\( N \) and the Hamiltonian. More precisely, the first perturbative
results exhibit the following features : 

1) \( N=A^{\dag }_{0}(T)A_{0}(T) \) is independent of \( t_{0} \), \( t_{1} \),
\( t_{2} \)..., i.e. of $T$. 

2)The \char`\"{}nonhomogeneous\char`\"{} parts of \( A_{n}(T) \) depend on
\( T \) through the basic operators \( A_{0}(T) \) and \( A^{\dag }_{0}(T) \).
The same is true for the \char`\"{}homogeneous\char`\"{} parts \( C_{n}(T_{1})\exp(-it_{0}) \)
which, after determination of \( C_{n}(T_{1}) \), can be recast in the form
of functions of \( A_{0}(T) \) and \( A^{\dag }_{0}(T) \) only. 

3) The operators \( H \) and \( N \) commute. 

If these features persist at all orders, then (putting aside any consideration
of convergence) one should obtain in the limit :

\begin{equation}
\label{29}
A(T,g)=A_{0}(T)+\sum _{n=1}^{\infty }g^{n}F_{n}(A_{0}(T),A^{\dag }_{0}(T))
\end{equation}
  where : 
\begin{equation}
\label{30}
A_{0}(T)=A_{0}\exp(-i(t_{0}+\sum _{n=1}^{\infty }f_{n}(N)t_{n}))
\end{equation}
 together with \( [A_{0},A^{\dag }_{0}]=1 \), and where the \( F_{n} \)'s
are some polynomial functions of \( A_{0} \) and \( A^{\dag }_{0} \) while
the \( f_{n} \)'s are some polynomial functions of \( N \). We will show below
that the resonance -free solutions of the perturbative multitime
equations of motion do exist indeed, and have the general form (29)-(30).
This means, in particular, that no obstructions are encountered in determining
the integration \char`\"{}constants\char`\"{} \( C_{n}(T_{1}) \) and giving
them the appropriate form. Eqs (29) and (30) then yield :\\

\begin{equation}
\label{31}
a(t,g)=a_{0}(t,g)+\sum _{n=1}^{\infty }g^{n}F_{n}(a_{0}(t,g),a^{\dag }_{0}(t,g))
\end{equation}
 where :
 
\begin{equation}
\label{32}
a_{0}(t,g)=A_{0}\exp(-i(1+\sum _{n=1}^{\infty }g^{n}f_{n}(N))t)
\end{equation}
 and \( N=A^{\dag }_{0}A_{0} \) is a constant operator.

Actually, these facts result from the full equivalence between the iterative
process described in the previous section and the perturbative determination of 
an unitary transformation which brings the Hamiltonian to a diagonal form.\\

In order to prove this equivalence, let us consider the spectral decomposition
of the Hamiltonian : 
\[
H(a_0,a^{\dag }_0,g) \equiv 1/2+a^{\dag }_0 a_0+g(a_0+a^{\dag }_0)^{4}/4=\sum _{n=0}^{\infty }E_{n}(g)|n,g><n,g|,\]
 where \( \left\{ \mid n,g>\right\}  \) is the
orthonormal basis made of the "perturbed" eigenvalues of $H$ (for future convenience, $a$ is
 written here as $a_0$). We also introduce the \char`\"{}unperturbed\char`\"{}, orthonormal
Fock basis \( \left\{ \mid n>\right\}  \) induced by the operators $a_0$ and $a^{\dag}_0$,
 together with the unitary transformation which maps the
former onto the latter : 
\[
|n>=U(g)|n,g>(n=0,1,2...).\]

The unitary operator \( U(g) \) is determined up to a \( N \) dependent, arbitrary,
right phase factor, where \( N=a^{\dag }_{0}a_{0} \).

Then if we define \( H_{d} \) as \( H_{d}(a_0,a^{\dag }_0,g)=U^{\dag}(g)H(a_0,a^{\dag }_0,g)U(g) \),
we have :

\begin{equation}
\label{33}
H_{d}(a_0,a^{\dag }_0,g)=H(a(g),a^{\dag }(g),g),
\end{equation}
 where we denote by \( a(g) \) the annihilation operator transformed by \( U(g) \)
:

\begin{equation}
\label{34}
a(g)=U^{\dag}(g)a_{0}U(g).
\end{equation}
 We also have :

\[
H_{d}(a_0,a^{\dag }_0,g)=\sum ^{\infty }_{n=0}E_{n}(g)|n><n|.\]

In other words, the unitary transformations (34) of the dynamical variables
is that one which diagonalizes the Hamiltonian in the \( \left\{ |n>\right\}  \)
basis. The perturbative form of eqs (33) and (34) are \\
 
\begin{equation}
\label{35}
a(g)=a_{0}+\sum _{n=1}^{\infty }g^{n}a_{n}
\end{equation}
 and 
\begin{equation}
\label{36}
H_{d}(a_0,a^{\dag }_0,g)=\sum _{n=0}^{\infty }g^{k}H_{k}=\frac{1}{2}+a^{\dag }_{0}a_{0}+\sum _{n=1}^{\infty }g^{k}H_{k}\, \, ,
\end{equation}
 where the explicit expressions of the \( H_{k} \)'s in terms of the \( a_{n} \)'s
are obtained by substituing the perturbative form (35) in \( H(a(g),a^{\dag }(g),g) \)
, and expanding. For the QAO Hamiltonian we are interested in, those$H_k$'s are: \\

\[
H_{k}=\sum _{n=0}^{k}a^{\dag }_{m}a_{k-m}+\sum _{m,r,s,\ell \geq 0\atop m+r+s+\ell =k-1}q_{m}q_{r}q_{s}q_{l}/4\,, (k=1,2,....) \, ,\]
 where \( q_{m}=a_{m}+a^{\dag }_{m} \).

The operators \( a_{n} \) (\( n=1,2,3\ldots  \)) in (35) are determined recursively
as polynomial functions of \( a_{0} \) and \( a^{\dag }_{0} \) by requiring
that :\\

i) \( U(g \)) be unitary indeed, or equivalently (due to Von Neumann theorem
\cite{RS}) that the commutator of the \char`\"{}new\char`\"{} variables \( a(g) \)
and \( a^{\dag }(g) \) be canonical: \( [a(g),a^{\dag }(g)]=1,\forall g \),i.e.:
\\ 

\begin{equation}
\label{37}
\sum _{m=0}^{n}[a_{m},a^{\dag }_{n-m}]=\delta _{n,0},(n=0,1,2...).
\end{equation}

ii) \( H_{d} \) be diagonal indeed in the \( \left\{ \mid n>\right\}  \) basis
or, equivalently, that $[H_d,N]=0,\forall g $, i.e. :
\begin{equation}
\label{38}
[H_k,N]=0,(k=1,2,3..)
\end{equation}

Eqs (38) implies that all the \( H_{k} \)'s are functions of the operator \( N=a^{\dag }_{0}a_{0} \)
only. In particular these operators \( H_{k} \) commute between themselves. 

Evidently, the equations (37) and (38) {\bf must} admit solution for \( \left\{ a_{n}\right\}  \),
due to the mere existence of the unitary operator \( U(g) \) and its formal perturbative 
expansion. However
there is no uniqueness property because of the phase freedom in the mapping \( U(g \)). In our case of QAO with standard quartic interaction,
the \char`\"{}minimal\char`\"{} solution \( \left\{ a_{n}\right\}  \) is such
that each \( a_{n} \) is a polynomial of degree \( 2n+1 \) in \( a_{0} \)
and \( a^{\dag }_{0} \) with rational coefficients, and monomials of odd degrees
only : 
\[
a_{n}=\sum _{2 k+\ell=2 n +1\atop 1\leq l_{odd}\leq 2n+1 }
                      (x_{n,k,l}a_{0}^{l}N^{k}+y_{n,k,l}N^{k}a_{0}^{\dag l}).\]
      
Let us now define the \( T \) dependent operators \( a_{n}(T) \) by:

\begin{equation}
\label{39}
a_{n}(T)=\exp(i\sum _{k=0}^{\infty }H_{k}t_{k})a_{n}\exp(-i\sum _{k=0}^{\infty }H_{k}t_{k}),
(n=0,1,2,...)
\end{equation}
 where \( t_{k}=g^{k}t \). We claim that these operators obey exactly the canonical
 commutation relations
(12) and the differential equations (10) which serve previously to determine
the \( A_{n}(T) \)'s. 

Because of (37) , this is immediate for the relations (12). As for the equations
(10) , one first derives from (39) :

\[
D_{n-m}a_{m}(T)=i\, \exp(i\sum _{k=0}^{\infty }H_{k}t_{k})[H_{n-m}(\{a_{r}\})
,a_{m}]\exp(-i\sum _{k=0}^{\infty }H_{k}t_{k})\]
which, summing up, yields

\begin{equation}
\label{40}
\sum _{m=0}^{n}D_{n-m}a_{m}(T)=i\sum _{m=0}^{n}[H_{n-m}(\{a_{r}(T)\}),a_{m}(T)],
\end{equation}
 where the commutativity of the \( H_{k} \)'s has been used twice. On the other hand,
 using (7) together with (33) and (36) to express \( H(A(T,g),A^{\dag }(T,g),g) \)
in terms of the function \( H_{k} \), one readily finds that the equations
(10) read as well:

\begin{equation}
\label{41}
\sum _{m=0}^{n}D_{n-m}A_{m}(T)=i\sum _{m=0}^{n}[H_{n-m}(\{A_{r}(T)\}),A_{m}(T)],
\end{equation}
 identical to (40). This is actually true not only for the QAO but for a general interaction. 

Therefore, \( \{A_{n}(T)\} \) can be identified as one of the solutions \( \{a_{n}(T)\} \),
which establishes the equivalence of the two schemes, and hence the consistence
of the multitime method we used, together with the general validity of the
assertions 1) to 3) put forward at the begining of this section.

It is possible now to comment on  the \char`\"{}mass renormalisation\char`\"{} introduced
 by B\&B. Since \( H_{d} \) is a pure function of $N$, \( H_{d}=H(N) \), eq(39) for $a_0(T)|_{t_{j}=g^{j}t} = A_0(T)|_{t_{j}=g^{j}t} = a_0(t,g)$ reads:

\[
a_{0}(t,g)=\exp(+iH(N)t)a_{0}\exp(-iH(N)t),\]
 or, by using \( a_{0}N=(N+1)a_{0} \) :

\[
a_{0}(t,g)=a_{0}\exp(-i(H(N)-H(N-1))t).\]

Together with (28), this gives:
\[
a_{0}(t,g)=a_{0}\exp(-it(1+3gN-3g^{2}(7+17N^{2})/4) +O(g^3)).\]

The \char`\"{}renormalisation\char`\"{} phenomenon can be pinned down to the fact that
 \( H(N)-H(N-1) \)
is the trivial identity operator at the zeroth order, and becomes a true operator
for higher orders.

As mentioned at the begining, and apparent on eqs (29) and (30), the solution
\( A(T,g) \) constructed there corresponds to initial conditions depending
on \( A_{0} \) {\bf and} \( g \) . If one insists in having the perturbative solution
with prescribed \( g- \)independent initial condition : \( A(0,g)=a \), with
\( [a,a^{\dag }]=1 \) , this is easily achieved by a few additional manipulations.
Indeed, it is sufficient to invert order by order the relation :

\[
a=A_{0}+\sum _{n=1}^{\infty }g^{n}F_{n}(A_{0},A_{0}^{\dag }),\]
 (which is straightforward in spite of the non commutative algebra) to get :\\

\[
A_{0}=a+\sum _{n=1}^{\infty }g^{n}G_{n}(a,a^{\dag }),\]
 and to reinsert this expression for \( A_{0} \) in eqs (29) and (30), as well
as in \( N=A^{\dag }_{0}A_{0} \), truncated at the relevant order. Then, of course, 
the expression of \( A(T,g) \) in terms of \( a \) and \( a^{\dag } \)
has no longer the \char`\"{}simple\char`\"{} structure that it exhibits in terms
of \( A_{0} \) and \( A_{0}^{\dag } \).

To conclude this section, we wish to stress again that the arguments presented there
are quite general, not specific of the QAO. If one considers an Hamiltonian
which is the sum of an harmonic oscillator one and a "potential" represented by
a self-adjoint operator function of the position and the momentum, such an analysis
can be repeated. Actually, the equivalence between MST and unitary
transformation diagonalizing the Hamiltonian is likely to be a rather general feature.
In particular, the previous discussion can be extended in a rather straigthforward
way to systems with more than one degree of freedom. 

Furthermore, the equivalence between the multitime approach and the perturbative
construction of the relevant unitary transformation must have a classical counterpart.
In the classical framework, multitime expansions should appear as essentially
equivalent to the construction of appropriate canonical transformations, following
the Poincar\'{e} - Von Zeipel method \cite{Gold}, or some of its disguises. As a matter
 of fact, one can find indication of such a connection in the literature \cite{N,M}. This
 aspect of the question, which we have not touched upon in the present paper, might deserve a further study.

\section*{4.~Conclusion}

In this paper, we have used the anharmonic oscillator in the Heisenberg picture as a model for investigating the practicability of the Derivative Expansion Method, of common use  
in classical physics, within the quantum framework. This
method turns out to be successful in providing us with the perturbative expansion of the time dependent
 dynamical variables together with the energy levels, which we have derived explicitely up to the second order. We also have proved that this MST is equivalent to the perturbative 
construction of an unitary transformation diagonalizing
 the full Hamiltonian, leading to a step-by-step algorithm for the calculation of the
 previous quantities at any order, and thereby strengthening the status of the Multiple Scale 
Techniques in quantum mechanics.

\newpage

\section*{{\large Acknowledgements} }

We are very grateful to Roberto Kraenkel who attracted our attention
to this present topic and Miguel Manna for enlightening discussions on the perturbative
multiscale theory.

\newpage

\section*{Appendix}
 We give below, up to order 6:

i) the coefficients $a_n$ of the expansion (35) of the annihilator $a(g)$ in terms of
 $N=a^{\dag}_0 a_0$,

ii) the coefficients $E_k(n)$ of the expansion of the energy levels:
\[
E_{n}(g)=\frac{1}{2}+n+\sum _{n=1}^{\infty }g^{k}E_{k}(n).\]

Both have been computed by applying the algorithm described in section 3 (eqs (37) and (38)).

\vspace{0.25cm}
i) To be simpler, \( a \) and \( a^{\dag } \) stand for \( a_{0} \)
and \( a^{\dag }_{0} \).
\vspace{0.25cm}

\( a_{1}=(2a^{3}-a^{\dag 3}-6Na^{\dag })/4 \) ;

\vspace{0.25cm}

\( a_{2}=(-9a+120a^{3}+6a^{5}-120a^{3}N+27aN^{2}+84a^{\dag }-42a^{\dag 3}-4a^{\dag 5}+42Na^{\dag 3}+276N^{2}a^{\dag })/32\, ; \) 

\vspace{0.25cm}

\( a_{3}=(6092a^{3}+756a^{5}+8a^{7}-8844a^{3}N+4422a^{3}N^{2}-378a^{5}N+1278aN-1062aN^{3} \)

\hspace{0.5cm}\( -1708a^{\dag 3}-464a^{\dag 5}-6a^{\dag 7}-9282Na^{\dag}+2406Na^{\dag 3}+232Na^{\dag 5}-1203N^{2}a^{\dag 3} \)

\hspace{0.5cm}\( -9042N^{3}a^{\dag })/128 \) ;

\vspace{0.25cm}

\( a_{4}=(-200645a+1546416a^{3}+380868a^{5}+9264a^{7}+40a^{9}-2975280a^{3}N+2143296a^{3}N^{2} \)

\hspace{0.5cm}\( -714432a^{3}N^{3}-322896a^{5}N+80724a^{5}N^{2}-3088a^{7}N-500298aN^{2}+162755aN^{4} \)

\hspace{0.5cm}\( +506760a^{\dag }-358200a^{\dag 3}-221832a^{\dag 5}-6696a^{\dag 7}-32a^{\dag 9}+673392Na^{\dag 3}+186464Na^{\dag 5} \)

\hspace{0.5cm}\( +2232Na^{\dag 7}+3040992N^{2}a^{\dag }-472788N^{2}a^{\dag 3}-46616N^{2}a^{\dag 5}+157596N^{3}a^{\dag 3}+1365240N^{4}a^{\dag })/2048\, ; \)

\vspace{0.25cm} 

\( a_{5}=(116798776a^{3}+51228696a^{5}+2189520a^{7}+20832a^{9}+48a^{11}-266946576a^{3}N \)

\hspace{0.5cm}\( +255315936a^{3}N^{2}-121842648a^{3}N^{3}+30460662a^{3}N^{4}-58614828a^{5}N+24750360a^{5}N^{2} \)

\hspace{0.5cm}\( -4125060a^{5}N^{3}-1300128a^{7}N+216688a^{7}N^{2}-5208a^{9}N+52602092aN+41073824aN^{3} \)

\hspace{0.5cm}\( -6417388aN^{5}-21539684a^{\dag 3}-28787584a^{\dag 5}-1542096a^{\dag 7}-16320a^{\dag 9} \)

\hspace{0.5cm}\( -40a^{\dag {11}}-121625250Na^{\dag }+50282976Na^{\dag 3}+32551232Na^{\dag 5}+912600Na^{\dag 7} \)

\hspace{0.5cm}\( +4080Na^{\dag 9}-47389884N^{2}a^{\dag 3}-13618080N^{2}a^{\dag 5}-152100N^{2}a^{\dag 7}-219914676N^{3}a^{\dag } \)

\hspace{0.5cm}\( +22248396N^{3}a^{\dag3 }+2269680N^{3}a^{\dag 5}-5562099N^{4}a^{\dag 3}-55675938N^{5}a^{\dag })/8192\, ; \)

\vspace{0.25cm}

\( a_{6}=(-2649077789a+19854323040a^{3}+14799326898a^{5}+1000498176a^{7}+15874840a^{9} \)

\hspace{0.5cm}\( +78720a^{11}+112a^{13}-15744a^{11}N-52410470592a^{3}N+59605775856a^{3}N^{2}-37821182832a^{3}N^{3} \)

\hspace{0.5cm}\( +13464443160a^{3}N^{4}-2692888632a^{3}N^{5}-20808622800a^{5}N+11852140500a^{5}N^{2} \)

\hspace{0.5cm}\( -3324992400a^{5}N^{3}+415624050a^{5}N^{4}-817924896a^{7}N+242212752a^{7}N^{2}-26912528a^{7}N^{3} \)

\hspace{0.5cm}\( -7253184a^{9}N +906648a^{9}N^{2}-16271788323aN^{2}-6097875991aN^{4}+521267535aN^{6} \)

\hspace{0.5cm}\( +4255953324a^{\dag }-2581523304a^{\dag 3}-8101045372a^{\dag 5}-691648560a^{\dag 7}-12242240a^{\dag 9} \)

\hspace{0.5cm}\( -64720a^{\dag {11}}-96a^{\dag {13}}+12944Na^{\dag {11}}+7571823000Na^{\dag 3}+11245609120Na^{\dag 5} \)

\hspace{0.5cm}\( +562441728Na^{\dag 7}+5584128Na^{\dag 9}+35458238196N^{2}a^{\dag }-9129805056N^{2}a^{\dag 3} \)

\hspace{0.5cm}\( -6326409960N^{2}a^{\dag 5}-165946104N^{2}a^{\dag 7}-698016N^{2}a^{\dag 9}+5783860872N^{3}a^{\dag 3} \)

\hspace{0.5cm}\( +1757503840N^{3}a^{\dag 5}+18438456N^{3}a^{\dag 7}+29695249188N^{4}a^{\dag }-2055444390N^{4}a^{\dag 3} \)

\hspace{0.5cm}\( -219687980N^{4}a^{\dag 5}+411088878N^{5}a^{\dag 3}+4768483548N^{6}a^{\dag })/65536. \)

\vspace{0.25cm}
ii)
\vspace{0.25cm}

\( E_{1}(n)=3(1+2n+2n^{2})/4 \) ;

\vspace{0.25cm}

\( E_{2}(n)=-(1+2n)(21+17n+17n^{2})/8 \) ;

\vspace{0.25cm}

\( E_{3}(n)=3(111+347n+472n^{2}+250n^{3}+125n^{4})/16 \) ;

\vspace{0.25cm}

\( E_{4}(n)=-(1+2n)(30885+49927n+60616n^{2}+21378n^{3}+10689n^{4})/128 \) ;


\vspace{0.25cm}

\( E_{5}(n)=3(305577+1189893n+2060462n^{2}+1857870n^{3}+1220765n^{4}+350196n^{5} \) 

\hspace{1.cm} \( +116732n^{6})/256 \) ;

\vspace{0.25cm}

\( E_{6}(n)=-(1+2n)(65518401+146338895n+213172430n^{2}+139931868n^{3}+85627929n^{4} \)

\hspace{1.cm} \( +18794394n^{5}+6264798n^{6})/1024. \)

\vspace{0.25cm}

Several number theoretic properties of the \( E_{k}(n) \)'s are worth pointing out. First, 
all the coefficients $c_{kp}$ of $n^p$ in $E_k(n)$ are rational and positive, 
and the signs of the $E_k(n)$'s alternate, as it should be. Perhaps new are the following
observations: whereas the denominator in the expression of $E_k(n)$ is a power of 2,
 the numerator is always a multiple of 3 (for integer $n$). This peculiarity was already 
noticed by Bender and Wu \cite{BW} for the ground state ($n=0$). It thus turns out to hold for the excited levels too. Also, the sum of the numerators of the coefficients $c_{kp}$ in each
$E_k(n)$ is a multiple of 5. Finally, if one expresses the $E_k(n)$'s in terms of the 
variable $m=n+\frac{1}{2}$, one observes that they are even polynomials with positive coefficients (multiplied by $-m$ if $k$ is even). More than that, all the zeroes of these polynomials
are pure imaginary. This means that all the zeroes of \( E_{k}(n) \) lie on the line $n=-\frac{1}{2} +iy$ !

\end{document}